\documentclass[aps,prb,twocolumn,superscriptaddress,longbibliography]{revtex4-2}

\usepackage[colorlinks=true,citecolor=blue]{hyperref}
\usepackage{graphicx}
\usepackage{amsmath}
\usepackage{amssymb}

\DeclareGraphicsExtensions{.png,.jpg,.eps}
\usepackage{xcolor}

\begin{document}

\title{Neural network enhanced hybrid quantum many-body dynamical distributions}
\author{Rouven Koch}
\affiliation{Department of Applied Physics, Aalto University, 00076 Aalto, Espoo, Finland}
\author{Jose L. Lado}
\affiliation{Department of Applied Physics, Aalto University, 00076 Aalto, Espoo, Finland}

\date{\today}

\begin{abstract}
Computing dynamical distributions in quantum many-body systems represents one of the paradigmatic open problems in theoretical condensed matter physics. Despite the existence of different techniques both in real-time and frequency space, computational limitations often dramatically constrain the physical regimes in which quantum many-body dynamics can be efficiently solved. Here we show that the combination of machine learning methods and complementary many-body tensor network techniques substantially decreases the computational cost of quantum many-body dynamics. We demonstrate that combining kernel polynomial techniques and real-time evolution, together with deep neural networks, allows to compute dynamical quantities faithfully. Focusing on many-body dynamical distributions, we show that this hybrid neural-network many-body algorithm, trained with single-particle data only, can efficiently extrapolate dynamics for many-body systems without prior knowledge. Importantly, this algorithm is shown to be substantially resilient to numerical noise, a feature of major importance when using this algorithm together with noisy many-body methods. Ultimately, our results provide a starting point towards  neural-network powered algorithms to support a variety of quantum many-body dynamical methods, that could potentially solve computationally expensive many-body systems in a more efficient manner.
\end{abstract}

\maketitle

\section{Introduction}
The dynamical and spectral properties of quantum many-body systems represents one of the 
central directions of modern quantum many-body physics. In particular, the development
of powerful methods such as auxiliary field
quantum Monte Carlo\cite{PhysRevD.24.2278,Sugiyama1986,2020arXiv201211914A}, 
tensor networks\cite{PhysRevLett.69.2863,Schollwck2011,2020arXiv200714822F,PhysRevLett.126.170603} 
and neural network quantum
states\cite{Carleo2017, PhysRevB.97.195136,choo2018symmetries,2021arXiv210305017V,PhysRevB.100.245123,PhysRevB.98.104426,PhysRevX.8.011006,PhysRevB.97.085104} have shed light onto the properties of paradigmatic quantum many-problems
that just a few years ago well drastically beyond conventional
methods\cite{PhysRevX.5.041041,Carleo2019,PhysRevLett.126.170603}. While the previous methods
have been shown highly successful for tacking static properties
of many-body systems, time-dependent and dynamical properties represent 
still a
critical open problem\cite{PhysRevLett.93.076401,PhysRevLett.93.040502,Silvi2019,PhysRevX.7.041015,PhysRevLett.122.250502,PhysRevLett.122.250501,PhysRevLett.122.250503,2021arXiv210301240V}. 
Spectral properties of many-body systems represent a central
issue in several novel directions of quantum-many body
physics, including non-equilibrium dynamics\cite{PhysRevLett.116.250401,RevModPhys.89.011004}, many-body
localization,\cite{RevModPhys.91.021001} 
many-body time crystals\cite{PhysRevLett.109.160401,PhysRevLett.117.090402,Sacha2017}
and dynamical phase transitions\cite{PhysRevLett.112.217204,Heyl2018}.
Interestingly, although a variety of methods to compute dynamical
properties exist that are affected in different forms by numerical
limitations and noise\cite{PhysRevLett.93.040502,PhysRevLett.93.076401,PhysRevLett.93.207205,PhysRevB.91.165112,PhysRevB.72.020404,PhysRevResearch.1.033009,PhysRevB.83.195115,2021arXiv210308804H,2019arXiv191208831L,PhysRevLett.125.100503,2021arXiv210104579B}, combining simultaneously this methods is a highly non-trivial
problem. 

\begin{figure}[t]
    \includegraphics[width=.93\columnwidth]{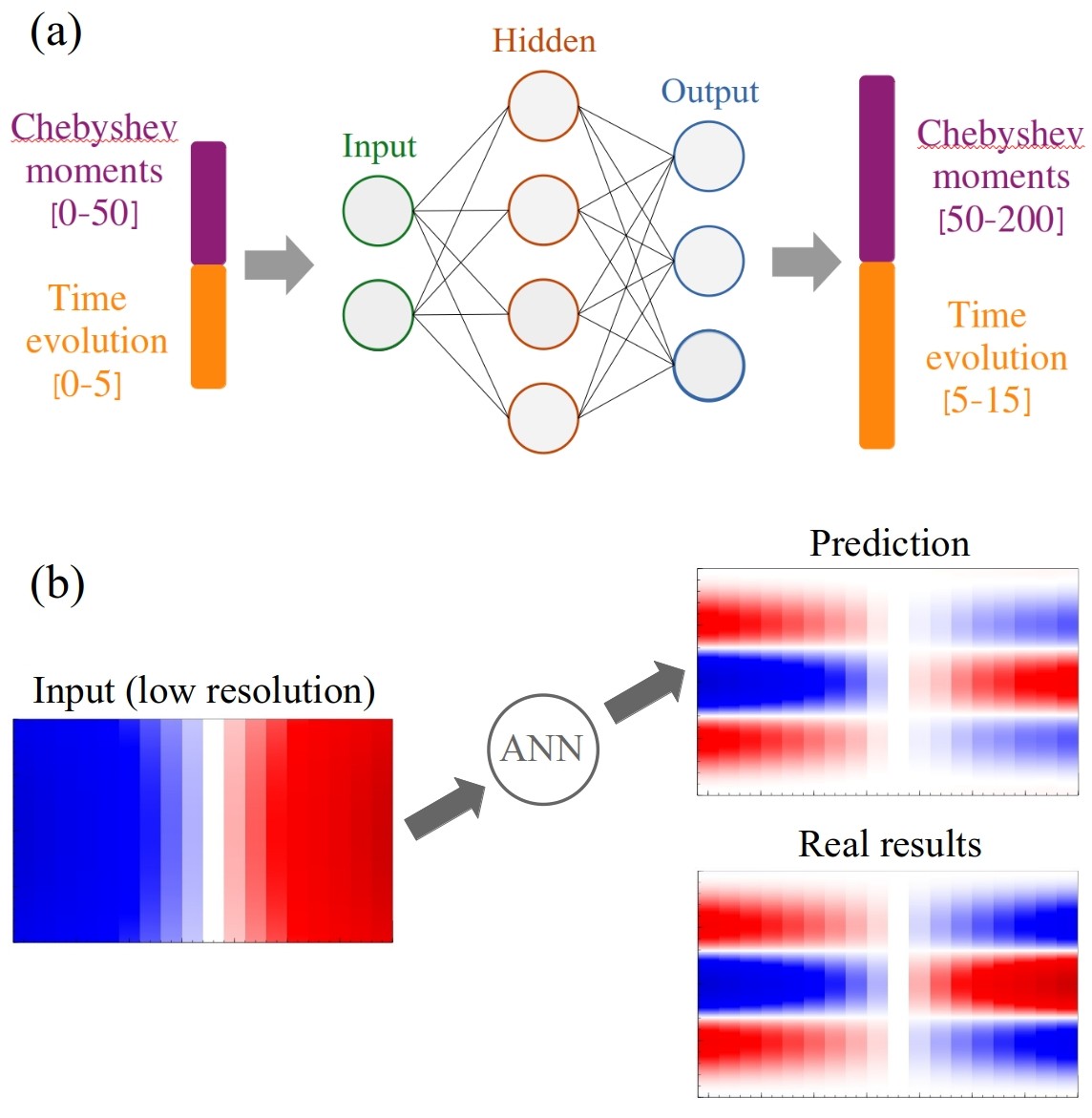}
    \caption{(a) Sketch of the hybrid neural-network many-body algorithm. The input of the ANN consists of Chebyshev moments 
    and time evolution,
    predicting higher-order moments
    and longer time evolution.
    The network is trained with single-particle, random-generated data and is able to make predictions of many-body systems. Panel (b)
    shows the comparison between the input low-quality data taken as input of the algorithm, and the output high-quality prediction
    together with
    its comparison with the real high quality data.
}
        \label{fig:ANN}
\end{figure}

Deep learning\cite{Goodfellow-et-al-2016, lecun2015deep} has risen as
a disruptive method for different fields of physics, including condensed matter physics\cite{carrasquilla2020machine,RevModPhys.91.045002}. Examples of these successful applications are neural network quantum states\cite{Carleo2017, PhysRevB.97.195136,choo2018symmetries,2021arXiv210305017V,PhysRevB.100.245123,PhysRevB.98.104426,PhysRevX.8.011006,PhysRevB.97.085104}, the detection of phases of matter\cite{carrasquilla2017machine,PhysRevB.97.115453,PhysRevB.99.245120,Greplova2020,vanNieuwenburg2017,RodriguezNieva2019,huembeli2019automated},
and machine learning strategies quantum control\cite{PhysRevResearch.1.033092,PhysRevApplied.13.054019}.
interestingly, the integration of machine learning methods with conventional many-body formalism
has the potential of drastically extending the applicability of typical methods, and even overcoming
conventional computational limitations\cite{PhysRevB.95.035105,Liu2019,PhysRevLett.124.020503,Fournier2020,PhysRevB.98.245101}.
However, up to date,
specific quantum many-body
are usually performed with
a single method, and potential synergies of combining 
them through in a non-trivial way several
of them have remained
relatively unexplored.

Here we demonstrate that machine learning methods combined with conventional dynamical-many body
techniques allows computing spectral properties dramatically increased accuracy.
In particular, we developed an algorithm combining results of the Kernel Polynomial Method (KPM)\cite{weisse2006kernel} and time evolution (TE) to calculate dynamical spectral distributions of single-particle and many-body systems. This hybrid Neural Network algorithm is able to drastically enhance the spectral properties of quantum many-body systems without being trained solely in single-particle data. We further demonstrate that our neural-network assisted
is extremely robust, being resilient to up to 10\% noise, making it ideal for real applications with state of the art methods showing numerical noise.
Our manuscript is organized as follows: in Section \ref{sec:methods} we present the details of our time-frequency algorithm,
in Section \ref{sec:single} we show its application to a family of single particle problems,
in Section \ref{sec:many} we show its extension to many-body systems without prior knowledge,
finally in section \ref{sec:con} we summarize our conclusions.

\section{Methods}
\label{sec:methods}

The method we put forward in our manuscript combines two well established techniques, kernel polynomial expansions and time evolution,
to compute the dynamical spectral properties of many-body system powered by a neural-network algorithm. The basic idea of our proposal
is shown in Fig. \ref{fig:ANN}, when the machine learning algorithm is trained, initial many-body data of the distribution is taken
as input, and the neural network returns as output a greatly enhanced many-body distribution. The many-body algorithm is solely
training in single-particle data, much cheaper to generate, and as we will show proves to successfully enhance many-body data. We
now elaborate on the different parts of the full method.

\subsection{The Kernel Polynomial Method and Chebyshev Expansion}
Quantity of critical interest in a variety of single particle and many-body systems can be written in the form

\begin{equation}
f(x)  =
        \langle \psi| A \delta (x - H) B | \psi \rangle 
        =
            \langle \alpha | \delta (x - H) | \beta \rangle 
\end{equation}
with $A,B,H$ operators and $\psi$ a certain state. In particular, dynamical spectral function such as dynamical spin structure
factors in spin systems, electronic many-body density of states and many-body magnetization distributions
can be represented in the previous form. From the computational point of view, a formal way of
computing such objects rely on full diagonalization of the operator $H$.
However,
the computation of the eigenvalues and eigenvectors of a matrix is for large systems,
in particular in a many-body system, is an almost unfeasible task. 
A greatly efficient alternative method is the Kernel Polynomial method and expansion in terms of Chebyshev polynomials \cite{weisse2006kernel}. Since the Chebyshev Polynomials are defined in the interval $I=(-1,1)$, a general expansion of $f:(-1,1) \rightarrow$ $R$ has the form
 \begin{equation}
 \label{eq:kpm1}
     f(x) = \alpha_0 + 2 \sum_{n=1}^\infty \alpha_n T_n(x) 
 \end{equation}
 where $\alpha_n=\langle f | T_n \rangle$ define the coefficients of the expansion and $T_n$ are Chebyshev Polynomials of order $n$. After rescaling the Hamiltonian and corresponding energies into the interval $(-1,1)$, the Chebyshev moments $\mu_n$ are defined as 
 \begin{equation}
     \mu_n = \langle \beta | T_n(H) | \alpha \rangle
 \end{equation}
 where $|\alpha\rangle$ and $|\beta\rangle$ are states of the system. Expansion of Eq. \ref{eq:kpm1} to finite order leads to so-called Gibbs oscillations. To avoid/minimize these oscillations a convolution of f(x) with a kernel is applied. In this paper, we are using the Jackson Kernel to reduce oscillations\cite{Jackson1912}.
 
 We now highlight the data used for training the machine learning algorithm.
 We will take of the density of states (DOS) of a single-particle system
 with eigenenergies $\epsilon_k$. The DOS of a given single-particle
 Hamiltonian can be written as
 \begin{equation}
     \rho(\omega) = \sum \delta(\omega-\epsilon_k)
 \end{equation}
 which can be calculated with the Chebyshev moments
 \begin{equation}
     \mu_n = \int_{-1}^{1} \rho(\omega)T_n(\omega)d\omega = \text{Tr}\left[T_n(H)\right]
 \end{equation}
 that can be computed with the recursion relations mentioned above.
 
 We now elaborate on the data that we will used for the many-body model, for which
 the machine learning algorithm trained in single particle
 data will be used. While we could use a many-body formulation of the electronic
 density of states, this choice could be considered trivial to test the generality
 of the algorithm, and therefore we will take a quantity that has no
 analogy in a single-particle framework.
 In a many-body spin system, in order to take the most different case possible with
 respect to the single-particle training,
 we will focus on many-body magnetization distributions, that lack an analogy for single particle models
 and are defined as
 
 \begin{equation}
     f(m) = \langle \Omega | \delta \left (m - \frac{1}{L} \sum_n S_n^z \right ) | \Omega \rangle
 \end{equation}
 where $S_n^z$ are the local spin operators and $|\Omega \rangle$ is the many-body ground state.
 These magnetization distributions can be computed within the matrix-product state
 formalism\cite{PhysRevLett.69.2863,Schollwck2011,2020arXiv200714822F},
 computing the many-body ground state implementing the
 Chebyshev expansion within the tensor-network algorithm.\cite{PhysRevB.83.195115,PhysRevB.90.115124,PhysRevB.90.045144,PhysRevResearch.3.013095,PhysRevResearch.2.023347,PhysRevResearch.3.013265,PhysRevResearch.3.023002}.
 
 \subsection{Time Evolution}
 Another approach to calculate the many-body spectral function
 or the density of states of single-particle systems
 on time evolution. For the training of the network, we will use time-evolution of the single-particle models
 under the single-particle Hamiltonian.
 Focusing first on the case of a single particle system,
 the time evolution of a single-particle system described by the Hamiltonian $H$ is given by
 \begin{equation}
 \label{eq:TE}
     f(t) = \langle \alpha | \texttt{exp}(-iHt) | \alpha \rangle \hspace{2mm} ,
 \end{equation}
 where $|\alpha \rangle$ is for example the state representing the lattice site where an electron is injected. In this case, $f(t)$ is the probability to find the electron at that lattice site at time $t$. The (local) DOS $\rho(\omega)$ is obtained by Fourier transformation of the function $f(t)$. 
 
 For the many-body case, we will use as input data time-evolution under the magnetization
 operator, defined as
  \begin{equation}
     f(t) = \langle \Omega | \texttt{exp}\left (-\frac{i}{L}\sum_n S_n^z \right ) | \Omega \rangle
 \end{equation}
where the Fourier transformation leads to the magnetization distribution of the system. 
 
\subsection{Linear Predictions}
While both time-evolution and Chebyshev expansion can be arbitrarily accurate, finite truncations either in time or order of moment leads to a finite spectral resolution.
In order to reduce these costs, prediction methods from linear predictions have been developed,
that formally attempt to extrapolate the behavior of moments or time-evolution. We will demonstrate that our neural-network method
drastically outperforms these linear prediction methods, and therefore for the sake of completeness we briefly introduce them here.
One of the most commonly used method for time-series predictions are linear-prediction models, e.g., autoregressive models~\cite{hyndman2018forecasting}. In the auto-regression model, the quantity of interest (here the next time steps of the time evolution or the Chebyshev moments) is predicted by a linear combination of $p$ previous data points. The model of order $p$ to predict the value $X_t$ at time step t can be written as
\begin{equation}
\label{eq:autoregression}
    X_t = c + \phi_1 X_{t-1} + \cdot\cdot\cdot + \phi_p X_{t-p} + \epsilon_t \hspace{2mm},
\end{equation}
where $c$ is a (time-independent) constant, $\epsilon_t$ white noise, and $\phi$ the $p$ parameters which define the model. The considered previous time steps, also called the lags, are used to include autocorrelations among adjacent data points. Equation \ref{eq:autoregression} allows to forecast an arbitrary number of time steps by using the predicted value as input for the next forecast. However, the accuracy (or confidence interval) decreases with every additional time step. Therefore, we are investigating more powerful prediction methods, which leads to the next section about Artificial Neural Networks (ANNs).
%
\subsection{Artificial Neural Networks}
The field of Machine Learning, starting from simple linear regressions up to Deep Neural Networks~\cite{Goodfellow-et-al-2016}, opens new opportunities and insights for the field of physics. In this paper, we illustrate the predictive power of fully-connected Deep Neural Networks and compare them with linear-prediction autoregression model. Deep neural networks consists of an input, output, and several hidden layers. The "neurons" of each layer are fully connected between neighbouring layers. The connections, the so-called weights, determine the outcome and predictions of the ANN and are the parameters which have to be tuned during the training process.
Using non-linear activation functions for the neurons in each layer allows the Neural network to represent a highly non-linear function. This function predicts the outcome of the network for an array of input values. In order to train the network, we are using the supervised learning
based on back-propagation., The neural network is trained with input data, i.e., in our case an input array consisting of a limited amount of Chebyshev moments (and time steps of the TE) and the output is an array of higher-order moments (and more time steps). The update of the weights happens during the training process via minimizing the loss function, in this case the mean squared error, and backpropagation. 
We use stochastic gradient decent and we are using the Adam optimizer~\cite{kingma2017adam}.

\section{Single-particle systems}
\label{sec:single}
\begin{figure}[t]
    \includegraphics[width=.8\columnwidth]{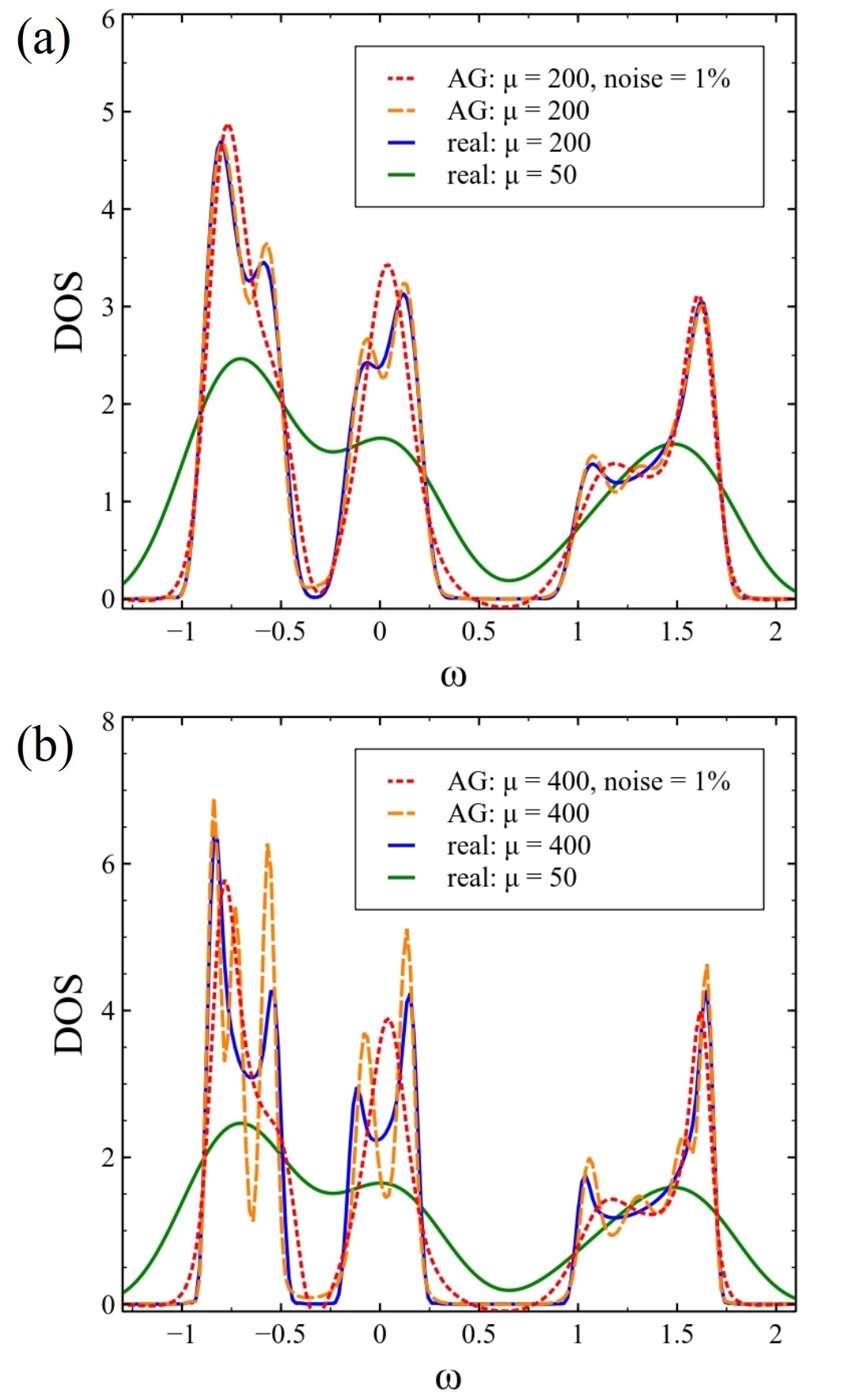}
    \caption{Linear predictions (autoregressive model) of the DOS of a random-generated tight-binding model with 1$\%$ (red) and without noise (orange) in the Chebyshev moments. The predictions up to 200 moments are shown in (a) and up to 400 in (b), respectively. The real distributions for 50 and 200 (400) moments are shown in green and blue.}
        \label{fig:results_complex_tb_linear}
\end{figure}

We now move on to consider the training data of our algorithm, single-particle data of a Hamiltonian,
which in particular is implemented as a wide family of one-dimensional tight binding models.

\subsection{Single-particle model}

The physical quantity of interest in case of the tight-binding system is the density of states (DOS), which can either be computed using the Kernel Polynomial method and the Chebyshev expansion or with the Fourier transformation of the time evolution. For more complex systems, e.g. many-body systems, the computation of higher-order moments and a long time evolution is very demanding. The higher the order of the moments used to calculate the DOS, the more details and features are captured and to obtain an accurate low-frequency DOS one needs a very long time-evolution. A common approach to obtain higher-order moments or a longer time evolution without an explicit calculation is to use linear prediction, e.g. an autoregressive model. With given a small amount of Chebyshev moments (and only the first time steps) the model can extrapolate the following moments (and time evolution). In this work, we present an alternative approach using an ANN algorithm to predict these quantities. The idea of the algorithm is to train an ANN to predict higher-order moments (and longer time evolution) given a limited amount of input moments (and time-steps), similar to the linear-prediction models. Our algorithm is capable of combining both approaches, i.e. forecasting higher-order moments and a longer time evolution at the same time.
\begin{figure}[t]
    \includegraphics[width=.8\columnwidth]{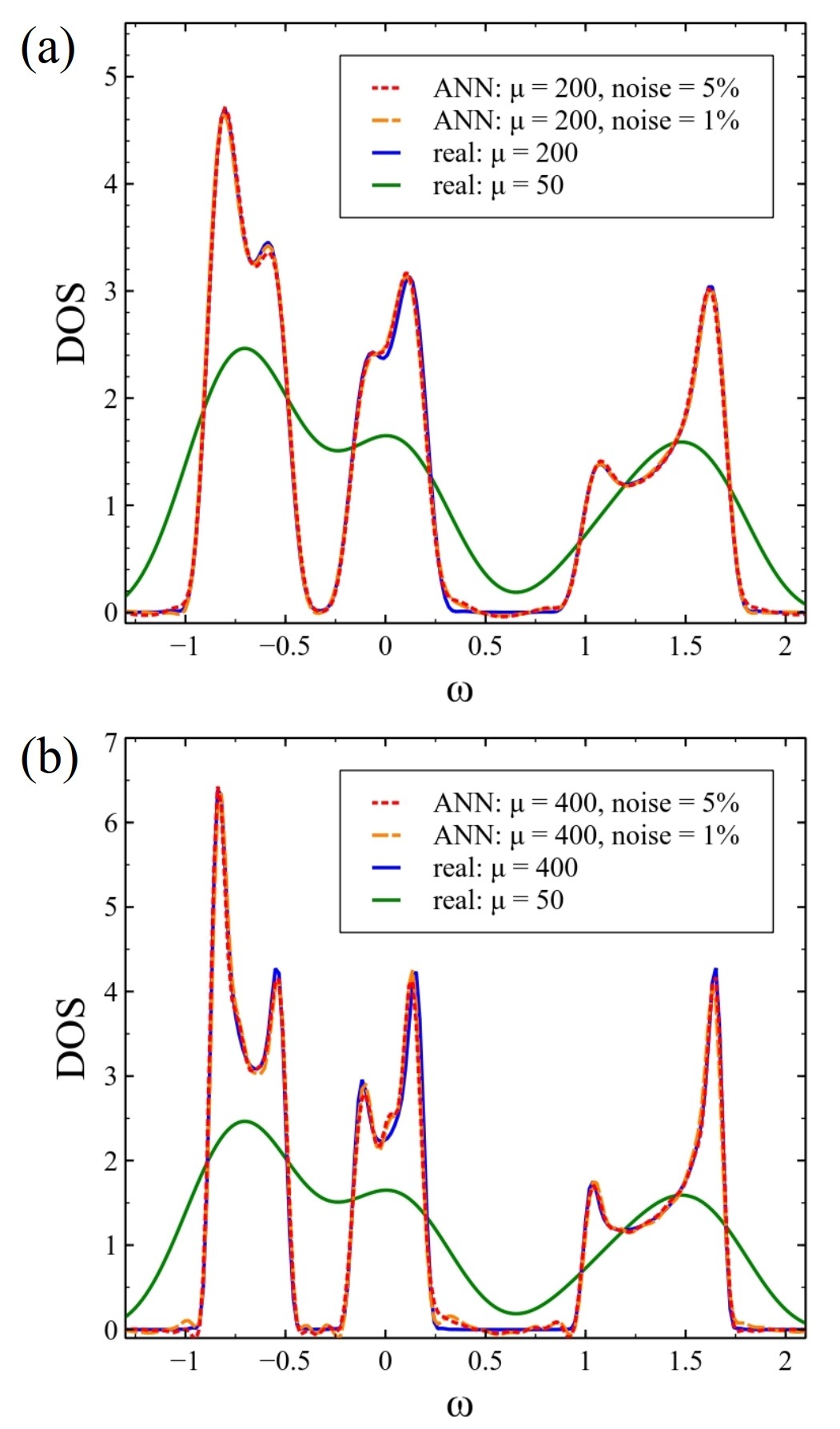}
    \caption{DOS for a random-generated tight-binding system out of the test set, calculated with 50 (green) and 200 / 400 (blue) Chebyshev moments. Predictions of the ANN with 1$\%$ (orange) and 5$\%$ (red) of noise for 200 and 400 moments are shown in (a) and (b).}
        \label{fig:results_tb_complex}
\end{figure}
We define a parametric family of spinless one-dimensional Hamiltonians of the form
\begin{equation}
    H_{TB} = \sum_{\langle i,j \rangle} t_{ij} c_i^\dagger c_j + \sum_{i} v_i c_i^\dagger c_i
    + \text{H.c.}
\end{equation}
where $t_{ij}$ is the hopping amplitude, $v_i$ the onsite energy, and $c^\dagger$ $(c)$ the electron creation (annihilation) operators.

\subsection{Neural-network versus autoregression}

Here we first show how the use of neural-network prediction provides
a greatly stable algorithm, overcoming limitations of autoregressive models.
For the sake of simplicity, the neural-network algorithm is trained only
with Chebyshev moments, and the combination with time evolution is
addressed in the next section.

As a starting point, we first consider how the density of states (DOS)
of these family of single particle models. 
For the neural-network,
the training is performed with single-particle 
models with translational symmetry
and three orbitals per unit cell.
We compute the DOS with the Chebyshev algorithm,
and we further compare with extrapolations performed with the autoregressive model.
Figure \ref{fig:results_complex_tb_linear} shows the DOS for an random-generated tight-binding systems. The higher-order moments are predicted with an autoregressive model, including a time window of 10 data points, given 50 moments as input and predictions up to order 200 and 400. This model achieves accurate results only in the case without noise up to 200 moments as shown in (a). For higher-order predictions (b) and noise values of about 1$\%$ the error of the autoregression model drastically increases and fails to predict the DOS (and, therefore, Chebyshev moments) with high accuracy - specific features and peaks of the original DOS are not captured. This shows the high sensitivity of linear predictions to noise,
and even in the noiseless severely limits the predictions (a maximum of a factor 4 in the noiseless limit). We now move on to show how neural-networks overcome
these limitations.

Deep neural-networks are in general, more robust when facing noise and can even benefit from it in order to make predictions more stable and reduce the probability of over-fitting. Figure~\ref{fig:results_tb_complex} shows the predicted DOS of the trained network, showing
a great agreement with the true distribution associated to the true moments. The Architecture of the ANN is described in the appendix. The model is trained with 10000 random generated tight-binding systems with different values for hopping amplitudes and onsite energies, given the first 50 moments as input vector and the next 350 moments as output during the training process. To show the robustness against noise, we induced 1$\%$ and 5$\%$ of noise in the Chebyshev moments. The results of Figure~\ref{fig:results_tb_complex} are that the predictions of the ANN algorithm are very accurate up to 200 moments (a and b) regardless of the induced noise. Even for predictions up to order 400 (factor 8) the ANN is able to predict the real DOS accurately, capturing all main features. In this case, some oscillations start appearing in comparison to the predictions up to 200 moments. Nonetheless, the enhancement of the input DOS (blue) and increased resolution is significant in all cases. This shows that deep neural networks a great advantage over the linear predictions, shown in Fig. \ref{fig:results_complex_tb_linear}.
\begin{figure}
    \includegraphics[width=1\columnwidth]{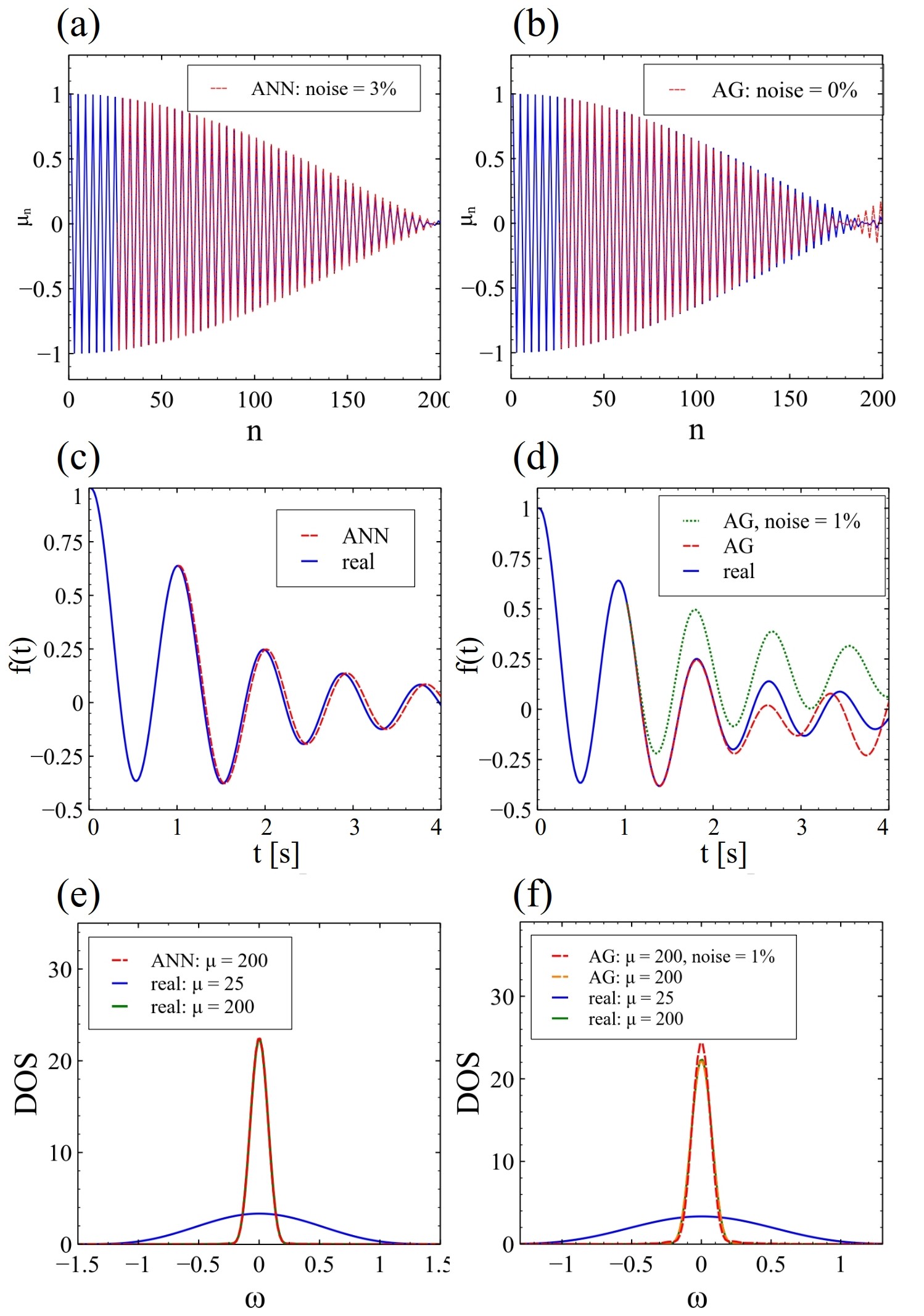}
    \caption{(Random-generated) tight-binding model similar to the expected many-body systems. The system is chosen out of the test-set of the ANN. The predictions (red) for the Chebyshev moments  of the ANN with 3$\%$ noise and AG model without noise are shown in (a) and (b). The corresponding time evolution predictions are shown in (c) and (d) including noise in the AG prediction (green). The corresponding DOS are shown in (e) and (f), comparing the input distribution (green) with the real (blue) and predicted one (red) including 200 instead of 25 moments.}
        \label{fig:results_tb_TE_ag}
\end{figure}

\subsection{Chebyshev and time-evolution neural-network architecture}

After showing the numerical robustness of the neural network, we now move on to considering
a more advanced neural network architecture in which both Chebyshev moments and time evolution
are included. The combination of these complementary allows to further test the reliability of the
machine learning algorithm, as the output and input contains both frequency and time information.
Overall, we find outstanding agreement for extrapolations of a factor 8 in the Chebyshev moments,
and with a strong reliance to numerical noise of up to 10\%.

We first comment on the training set. The training models consist of randomly generate 5000 systems, consisting of two sites per unit cell with different hopping amplitudes. The architecture of this model is described in the appendix and similar to the one for the many-body predictions. The predictions of the trained model are shown in Fig.~\ref{fig:results_tb_TE_ag} (a,c,e) and compared with linear predictions of the same system (b,d,f). Another difference to the previous model is that we are now predicting the Chebyshev moments (a) and time evolution (c) at the same time, i.e. including both in the same input array for the ANN. We use the first 25 moments and 50 time steps as input and calculate up to 200 moments and 200 time steps. The predictions shown in Fig.~\ref{fig:results_tb_TE_ag} (a) and (c) are very accurate, especially for the moments and the resulting DOS prediction (e) is almost indistinguishable from the original one. In the predictions of the moments, we induced noise of 3$\%$ in order to show the robustness of the ANN and also decrease the risk of overfitting. The averaged root mean squared error (RMSE) of the test set (500 random generated systems) for the predictions of the moments is 0.008 and for the time evolution 0.005. The enhancement of the resolution in comparison to the input DOS (blue) is significant and shows a pronounced peak. The combination of the moments with the time evolution are one of the main advantages of the linear prediction models. Fig.~\ref{fig:results_tb_TE_ag}(b) and Fig.~\ref{fig:results_tb_TE_ag}(d) shows the predictions of the autoregression model (without noise) for the Chebyshev moments and time evolution. The predictions of the moments are accurate up to order 170 and the time evolution without nose up to 115 time steps. The induced noise in the TE leads to instabilities of the linear model and drastically decreases the accuracy. The same applies for noise in the Chebyshev moments which leads to a decreased accuracy in the prediction of the DOS, shown in Fig.~\ref{fig:results_tb_TE_ag}(f).

With the previous neural-network architecture, we now move on to extrapolate the many-body distributions
of a quantum many-body system.

\section{Many-body systems}
\label{sec:many}

We now move on to consider how our trained neural-network performs when
presented with new data, now originating from a many-body Hamiltonian. In particular,
in order to demonstrate the generality of the training, we will focus
on the many-body magnetization distribution of a spin-chain, a quantity
that has no analogy with the single-particle density of states used in the
training of the machine learning algorithm.

\subsection{Many-body model}
Here we will consider a many-body spin Hamiltonian, in the form of a $S=1/2$ one-dimensional many-body system
defined by the Hamiltonian

\begin{equation}
\label{eq:HS12}
    H = J \sum_n S_{n}^x S_{n+1}^x + S_{n}^y S_{n+1}^y + S_{n}^z S_{n+1}^z 
\end{equation}

First, it is worth to note that $S=1/2$ one dimensional models can be generically mapped
to interacting spinless fermionic models by means of a Jordan-Wigner transformation\cite{Giamarchi2003}. 
In this context,
spin-excitations in the many-body spin chain 
$\langle \Omega | S_n^+ \delta (\omega - H + E_0 ) S_n^-|\Omega \rangle$
are mapped to quasiparticle spectral functions in the interacting
fermionic model
$\langle \Omega | c^\dagger_n \delta (\omega - H + E_0 ) c_n |\Omega \rangle$.
In this regard, deploying the trained hybrid neural-network algorithm
to compute spin-excitations
of the interacting Heisenberg model would be a relatively trivial test of our algorithm.
In order to target a more non-trivial problem, here we will focus on the many-body magnetization distribution,
which cannot be mapped to a single-particle property, and therefore represents a more advanced test of the algorithm.

The physical quantity of interest is for the many-body systems the magnetization distribution,
defined as

\begin{equation}
    f(m) = \langle \Omega | \delta  (m - \hat M_z  ) | \Omega \rangle
\end{equation} 
with $\hat M_z = \frac{1}{L}\sum_n S_n^z $, with $L$ the number of sites
and $| \Omega \rangle$ the many-body ground state of the Hamiltonian $H$ of Eq. \ref{eq:HS12}.
This magnetization distribution provides reflects the spectral composition of the ground state in terms of eigenstates
of $\hat M_z$, in particular reflecting the many-body entanglement of the system\cite{PhysRevLett.100.165706,PhysRevE.67.056129,Gritsev2006}. As noted above, this quantity does not map
to a meaningful object in the single-particle limit, and therefore represents a non-trivial test for the hybrid neural-network algorithm.
The previous quantity is computed both via the Chebyshev expansion and time-evolution methods, and used as input for the neural network.
As a reference, we will also compare linear predictions with the results of the ANN algorithm for a many-body system.

\begin{figure} 
    \includegraphics[width=.75\columnwidth]{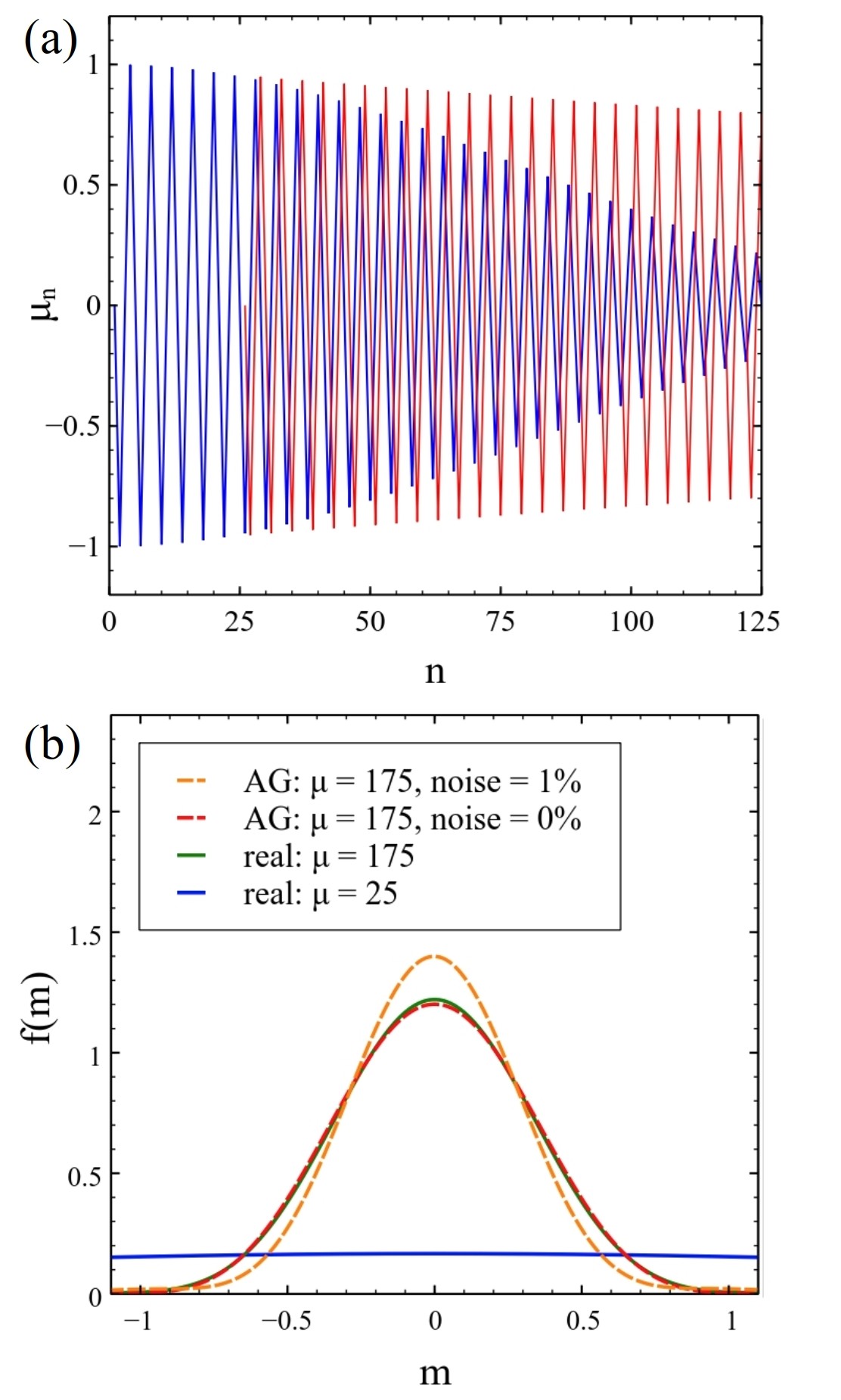}
    \caption{Linear predictions (red) of the autoregressive model of the Heisenberg S1/2 chain (30 sites) for the Chebyshev moments with 1$\%$ noise (a). The input for the model are the first 25 moments and the first 50 time steps for the time evolution. The resulting magnetization distribution without (red) and with 1$\%$ (orange) noise is shown in (b).}
        \label{fig:results_HS12_ag}
\end{figure}

\subsection{Autoregressive model}

We first address the performance of autoregressive models for extrapolating the many-body
magnetization distribution.
The results of the autoregressive model are shown in Fig. \ref{fig:results_HS12_ag} and lead to similar conclusions as in the single-particle case, where we studied the DOS in Fig. \ref{fig:results_complex_tb_linear}. The predictions of the Chebyshev moments with 1$\%$ noise are shown in (a). The induced noise in the moments of 1$\%$ or even less leads to instabilities in the model and decreases the accuracy enormously. This can also be seen in the corresponding magnetization and the loss in accuracy in (b).
\begin{figure}
    \includegraphics[width=\columnwidth]{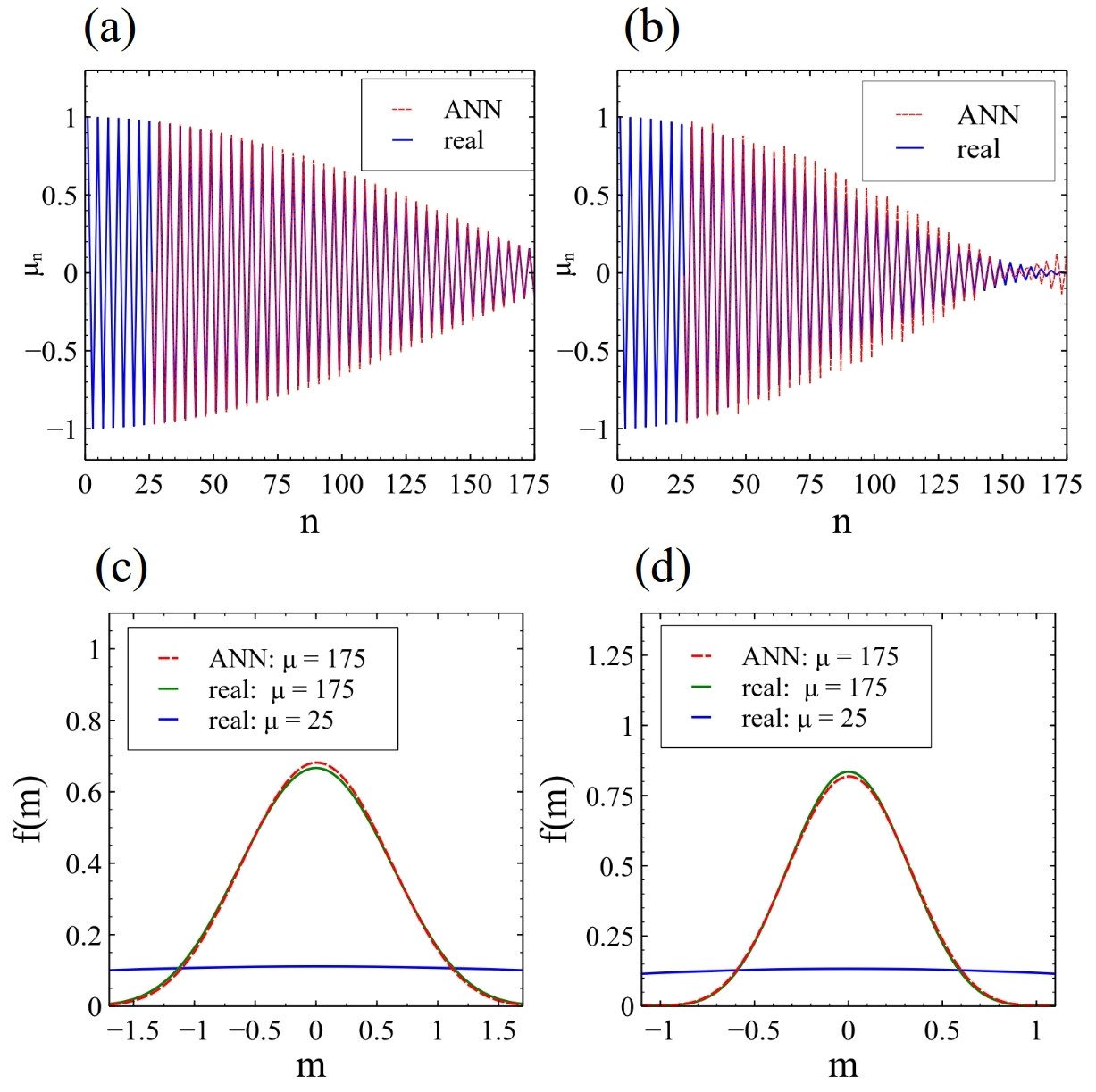}
    \caption{Predictions of the ANN for the (1d) Heisenberg S$1/2$ model. The Chebyshev moments and time evolution are predicted simultaneously with an input array consisting of the first 25 moments and 50 time steps. The predictions (red) of the trained ANN of the moments for system size $L=10$ and $L=30$ are shown in (a) and (b). The corresponding magnetization distribution in in (c) and (d). The magnetization distribution is calculated with the input moments (blue) in comparison with the real (green) and predicted (red) distribution for 175 moments.
    }
\label{fig:results_HS12_ANN}
\end{figure}

\subsection{Hybrid neural-network many-body algorithm}

After showing the sensitivity of autoregression to noise, we now move on to use our training hybrid neural-network
architecture in the many-body distribution problem.
The robustness to noise is one of the main advantages and motivations to develop an ANN algorithm to predict higher-order moments (and TE) and enable the calculation of the magnetization distribution of complex systems with higher resolution. The training data remains the same as for the single-particle system - random-generated tight-binding systems with a similar DOS as the magnetization distribution of the many-body systems. The ANN is, therefore, trained with purely single-particle physics and predictions are made for many-body systems without prior knowledge.

The predictions of our ANN algorithm for the one-dimensional Heisenberg $S=1/2$ model are shown in Fig. \ref{fig:results_HS12_ANN}. The algorithm predicts Chebyshev moments and the time evolution simultaneously given an input array of 25 moments and 50 time steps. The predictions of the moments above order 170 become less accurate, achieving an accurate prediction of factor 7. 
The magnetization distributions for the Heisenberg $S=1/2$ chains is shown in Fig. \ref{fig:results_HS12_ANN} (b). Shown are the input distribution using only 25 moments and the real distribution for 175 and the predictions of the ANN algorithm. The predictions show high accuracy and match the expected distributions well for both system sizes. They also show an enormous enhancement to the almost flat input distribution which does not show the pronounced peak. Therefore, the fluctuations in the moments in (b) do not have a significant impact on the distribution. From this follows, that the ANN algorithm is capable of enhancing the many-body distribution significantly. The resolution of the predictions is however limited up to a certain order of Chebyshev moments. In comparison with the AG magnetization distributions (Fig. \ref{fig:results_HS12_ag} (b)) the ANN algorithm is able to match the real distribution with higher accuracy. 

\begin{figure}
    \includegraphics[width=\columnwidth]{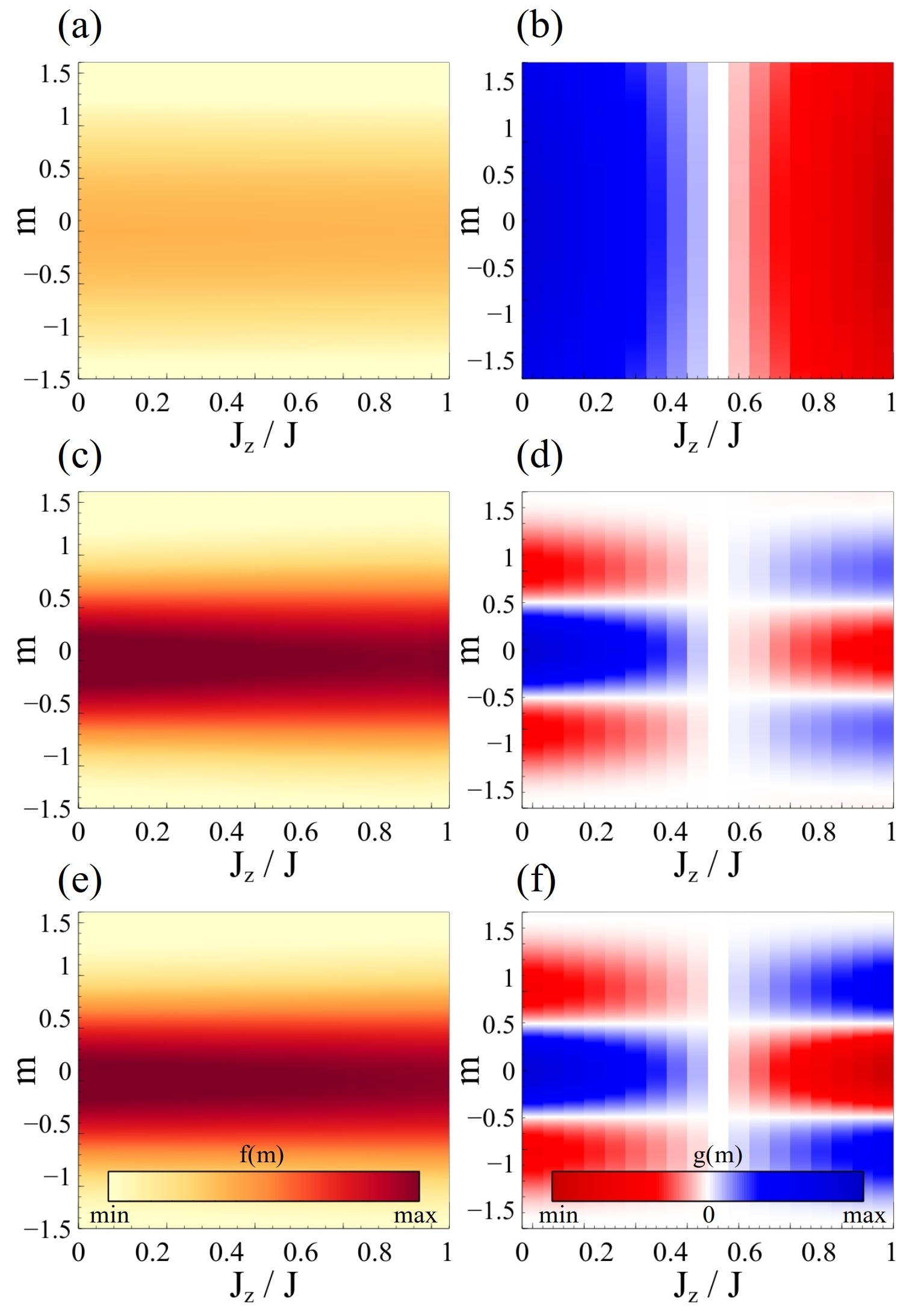}
    \caption{Magnetization distribution of the quantum many-body model
    of Eq. \ref{eq:xxz}. 
    Panels (a,b) show the low-quality distribution,
    (c,d) the predicted distribution by the NN, and
    (e,f) the true distribution.
    Panels (a,c,e) show the
    original distribution $f(m)$, and (b,d,f)
    background-subtracted distribution $g(m)$.
    It is observed that the predicted distribution (c,d)
    and the real one (e,f) show a substantial agreement.
    }
\label{fig:map}
\end{figure}

We now use our trained neural network algorithm to study the magnetization distribution of a generalized
many-body model, namely the XXZ $S=1/2$ model defined as
\begin{equation}
    H(J_z) = J \sum_n [S_{n}^x S_{n+1}^x + S_{n}^y S_{n+1}^y] + J_z \sum_n S_{n}^z S_{n+1}^z 
    \label{eq:xxz}
\end{equation}
For $J_z=1$, the previous model corresponds to the Heisenberg model studied previously.
The magnetization distribution in the $z-$ depends on the specific value of $J_z$. In the following
we show how the neural network successfully predicts the evolution of the magnetization distribution
in this generalized model. To highlight the differences for the different values of $J_z$, we will
now plot the difference of the distribution with respect to the
ground state for $J_z/J=0.5$, and normalized to the maximum value fo the distribution, namely
$f(m,J_z) = \langle \Omega_{J_z} | \delta (m-\hat M_z)| \Omega_{J_z}\rangle $,
where $| \Omega_{J_z}\rangle $ is the many-body ground state of $H(J_z)$. Furthermore,
to highlight that the neural-network captures even the fine changes in the distribution,
we define a background subtracted distribution of the form $g(m,J_z) = f(m,J_z) - f(m,J_z=0.5J)$,
that allows to observe the subtle changes in the distribution with respect to the one
for $J_z=0.5J$. We show in Fig.~\ref{fig:map} the distributions $f(m)$ (Fig. \ref{fig:map}(a,c,e))
and $g(m)$ (Fig.~\ref{fig:map}(b,d,f)). In particular,
Figs. \ref{fig:map}(a,b) show the low-quality distributions, that represent the input of the trained algorithm.
The predicted distribution by our neural-network is shown in Figs .\ref{fig:map}(c,d), which shows a great agreement
with the true high quality real distributions shown in Figs. \ref{fig:map}(e,f). The previous results
further demonstrate that our trained algorithm successfully predicts distributions of the generalized model
of Eq.~\ref{eq:xxz},
even capturing fine details of the distribution.

\section{Summary}
\label{sec:con}
We have shown that neural networks, in combination with standard methods for computing
spectral properties, provide a powerful framework to predict dynamics
with increased accuracy. In particular, taking as examples
time-evolution and Chebyshev methods, we have shown that reliable extrapolations
can be performed.
This was demonstrated first for a single-particle problem, where
our architecture greatly improved the spectral resolution of the density
of states. We then showed that the hybrid neural-network architecture
trained in single-particle data could be deployed to enhance the
many-body magnetization distribution of a purely many-body problem,
a problem drastically different from the training set of the algorithm.
Importantly, we have shown that our method is highly resilient to
noise and provides dramatically more stable predictions than conventional
linear regression methods.
These results can be further generalized to other methods of computing
spectral properties, including correction vector
methods and imaginary time-evolution.
Importantly, while our results have been
implemented with a tensor-network formalism,
an analogous procedure can be
carried out with generic quantum many-body methods.
Our results demonstrate that neural-network enhanced
many-body methods provide a numerically robust formalism
to compute many-body spectral properties.

\textbf{Acknowledgements}
We acknowledge the computational resources provided by
the Aalto Science-IT project,
and financial support from the
Academy of Finland Projects No.
331342 and No. 336243.
We thank C. Flindt and T. Kist
for fruitful discussions.

\section*{Appendix}

\appendix

\section{Deep Neural Network architectures}

The ANN architecture of the single-particle systems with only Chebyshev moments (see Fig. \ref{fig:results_tb_complex}) is defined as follows: an input layer of dimension 50, 3 hidden layers of dimensions 256, 512 and 512 with 15$\%$ of dropout \cite{srivastava2014dropout}, and an output layer of dimension 350. As activation function we are using the non-linear ReLu function. The test and training data consists of 10000 random generated tight-binding models (90/10 split) defined with a 3 site unit cell, with different onsite energies and hopping amplitudes. We choose the Adam optimizer and optimized the mean squared error. The training process includes 300 epochs with a batch size of 16.

The ANN architecture for the combined Chebyshev and time-evolution input consists of an input layer of dimension 75, three hidden layers of size 512, 1024 and 512 with 20$\%$ dropout to avoid over-fitting, and an output layer of dimension 525. As activation function we are using the non-linear ReLu function. The training data consists of 5000 random generated tight-binding systems, in this case the unit cell consists of 2 sites with different hopping amplitudes and zero onsite energies. The training process is done with a batch size of 50 for 512 epochs and a train-test-split of 90/10. In addition, we included 5$\%$ of noise in the Chebyshev moments and 1$\%$ in the time evolution.

\bibliography{references}{}

\end{document}